\documentclass[conference]{IEEEtran}
\IEEEoverridecommandlockouts
\usepackage{cite}
\usepackage{amsmath,amssymb,amsfonts}
\usepackage{algorithmic}
\usepackage{graphicx}
\usepackage{textcomp}
\usepackage{xcolor}
\usepackage{subfig}
\usepackage{booktabs}
\usepackage{import}
\usepackage{color}
\usepackage[citebordercolor=blue]{hyperref}
\usepackage[capitalise]{cleveref}
\usepackage{multirow}
\usepackage{array}
\usepackage[font=footnotesize]{caption}
\usepackage[inline]{enumitem}
\usepackage{makecell}
\usepackage{siunitx}
\usepackage{caption}
\usepackage{soul}
\sethlcolor{yellow}

\usepackage{tikz}
\newcommand\copyrighttext{%
  \footnotesize \textcopyright 2026 IEEE. Personal use of this material is permitted. Permission from IEEE must be obtained for all other uses, in any current or future media, including reprinting/republishing this material for advertising or promotional purposes, creating new collective works, for resale or redistribution to servers or lists, or reuse of any copyrighted component of this work in other works.}
\newcommand\copyrightnotice{%
\begin{tikzpicture}[remember picture,overlay]
\node[anchor=south,yshift=10pt] at (current page.south) 
  {\fbox{\parbox{\dimexpr\textwidth-\fboxsep-\fboxrule\relax}{\copyrighttext}}};
\end{tikzpicture}%
}

\begin{document}
\bstctlcite{BSTcontrol}
%
\title{Constellation-Level Power Allocation for LEO Space-Based Solar Power}
%
%
%
\author{\IEEEauthorblockN{
    Mustafa~Alhassan,~Amjad Iqbal~and~Peng~Hu \\
\IEEEauthorblockA{
    Advanced Network and Embedded Systems Lab (AEL)\\
    Dept. of Electrical and Computer Engineering, University of Manitoba, Winnipeg, Canada}
    alhass11@myumanitoba.ca,\{Amjad.Iqbal, Peng.Hu\}@umanitoba.ca
}
\thanks{We acknowledge the support provided by the Government of Canada, Natural Sciences and Engineering Research Council of Canada (NSERC), [funding reference number RGPIN-2022-03364], and Research Manitoba.}
}

\markboth{Journal of \LaTeX\ Class Files,~Vol.~14, No.~8, August~2015}%
{Shell \MakeLowercase{\textit{et al.}}: Bare Demo of IEEEtran.cls for IEEE Journals}

\maketitle
\copyrightnotice
\begin{abstract} Space-based solar power (SBSP) has recently gained renewed attention as an appealing technological advancement for providing continuous clean energy using space-based infrastructure. However, the potential of low-Earth orbit (LEO) satellite constellations for SBSP remains largely unexplored and lacks detailed simulation-based studies. In this paper, we introduce a novel LEO SBSP system model and conduct a 24-hour system-level simulation of a Walker 4$\times$5 LEO SBSP constellation at an altitude of 450\,km, beaming 2.45\,GHz microwave power to eight ground stations (GSs) under a greedy allocation policy. The model includes orbital propagation, eclipse cycles, the satellite power chain, Goubau--Brown beam coupling, ITU-R P.618 atmospheric attenuation, and onboard battery dynamics. The results confirm that the peak DC power delivered reaches 1.986\,MW, while the mean per-site delivery at the served GS ranged from 40 to 75\,kW. Two of the eight GSs received no service during the run, as their passes were consistently ranked lower under the greedy policy than competing links at the same step. The incident peak power density (PD) at the rectenna remained within the 3.35--5.72 W/m\textsuperscript{2} range, below the International Commission on Non-Ionizing Radiation Protection (ICNIRP) general-public exposure limit. For a 20-satellite Walker LEO at this altitude, realistic per-site delivery is 50--100 kW, and the rectenna should be sized to the operational incident PD of order 5 W/m\textsuperscript{2} rather than to a Geostationary Earth Orbit (GEO)-era 100 W/m\textsuperscript{2} rating.
\end{abstract}


\IEEEpeerreviewmaketitle

\section{Introduction}
Space-based solar power (SBSP) is a transformative vision for global energy sustainability by 2050, aiming to provide continuous, 24/7 clean energy to help achieve net-zero emissions targets. SBSP collects sunlight energy in orbit with photovoltaic (PV) panels and transmits it to ground-based rectennas via microwave/laser technologies. This concept was first introduced by Glaser in 1968~\cite{Glaser_1968_PowerFromTheSun}. SBSP places satellites in orbit, avoiding disruptions caused by clouds, weather, and the day-night cycle that affect terrestrial solar farms~\cite{kang_harnessing_2024}. Moreover, the time-averaged solar flux received by a satellite is several times higher than that available at Earth's surface~\cite{sasaki_microwave_2013}.
\par Most SBSP designs rely on a satellite in Geostationary Earth Orbit (GEO), where the satellite remains fixed relative to Earth, and a single large power station that simultaneously transmits energy to a fixed rectenna~\cite{marshall_investigation_2022,guo_leo_2023}. GEO designs require a large-scale aperture (kilometer-scale) and launch capabilities that are still years away~\cite{Rodgers_SBSP_2024}. According to~\cite{guo_leo_2023}, approximately one hundred low Earth orbit (LEO) satellites with smaller individual transmitters can match a single GEO station, and a four to five-station equatorial medium Earth orbit (MEO) constellation can match GEO economics~\cite{marshall_investigation_2022}. To date, very little practical work has been conducted on the real-world implementation of SBSP. For example, the Caltech MAPLE experiment demonstrated phased-array power transfer from a CubeSat to the ground in 2023~\cite{ayling_wireless_2024}. China plans to implement a long-term roadmap to develop SBSP, ranging from kilowatt-class LEO test articles to MW-class GEO stations~\cite{cheng_phased_2025}.

The problem of real-time resource allocation across a multi-satellite LEO SBSP constellation serving multiple ground sites has received limited attention in the literature. At LEO altitudes, each satellite is visible to a given station for only a few minutes per pass, and solar input drops to zero during eclipse. The constellation must therefore decide, at each timestep, which satellite--station pairs to activate, how much RF power to transmit on each link, and how much energy to reserve in the on-board batteries for the upcoming eclipse. Earlier system-level work has considered single-GEO load chasing against a single-city demand profile ~\cite{madonna_feasibility_2024}, but the multi-site case at LEO has not been addressed. 

To the best of our knowledge, this is the first study to schedule microwave power transfers across a LEO SBSP constellation serving multiple ground stations (GSs), subject to realistic orbital, atmospheric, and battery constraints. We present a 30-second step simulation of a Walker 4 $\times$ 5 layout at an altitude of 450 km, beaming to eight GSs using microwave technologies. We consider a 24-hour run under a greedy allocator policy, during which peak DC power delivered reached 1.986 MW, and mean per-site power delivery at the served stations was 40-75 kW, with incident power density (PD) well below the ICNIRP general-public limit. We find that realistic per-site delivery for a constellation of this size is 50--100 kW per station, and that the rectenna should be sized to the actual incident PD rather than to a GEO-era-rated value.
\begin{figure*}[!t]
\centering
\includegraphics[width=\textwidth]{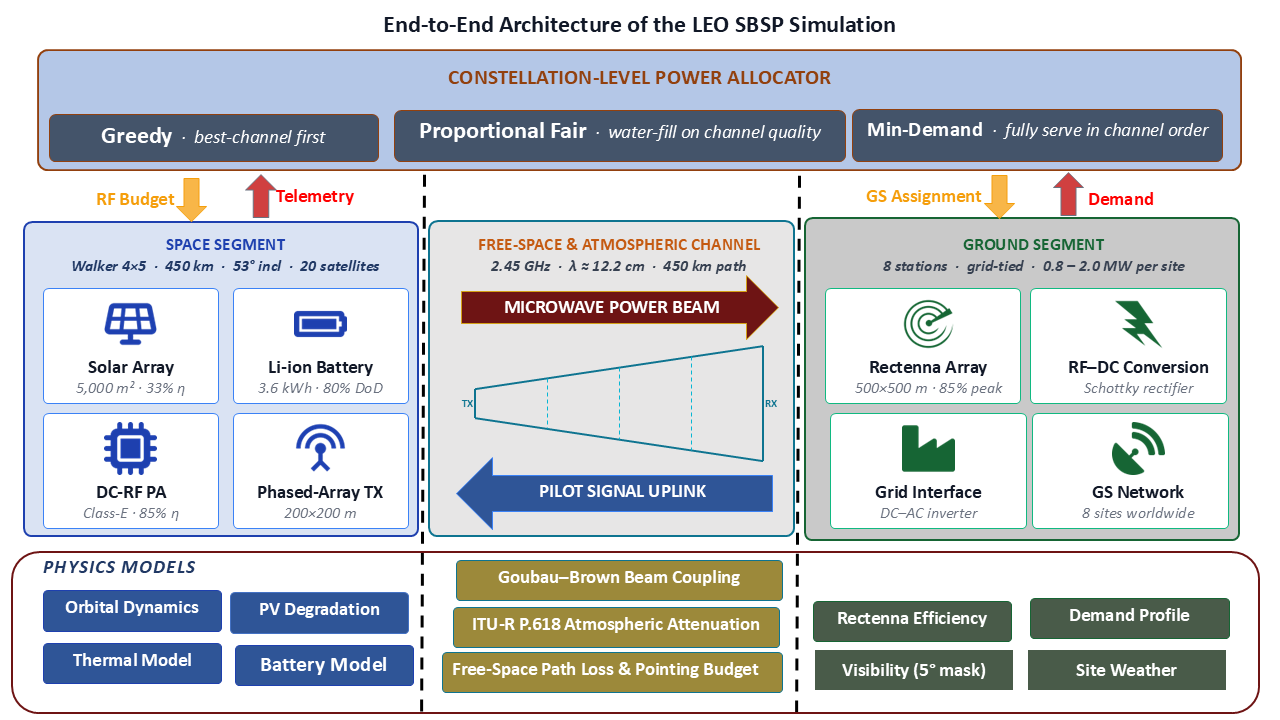}
\caption{End-to-end architecture of the LEO SBSP framework. The power allocator at the top decides, every 30 seconds, how much RF power each of the 20 satellites should send to each of the 8 GSs. It can use one of three policies: Greedy, Proportional Fair, or Min-Demand. No satellite may deliver more than 1.05 times a station's demand. Below the allocator, the system is split into three parts. The Space Segment generates power and transmits the beam. The Free-Space and Atmospheric Channel carries the beam through space and the atmosphere, where it is shaped by Goubau--Brown beam coupling and attenuated by rain and gases, as described in ITU-R~P.618. The Ground Segment receives the beam, converts it to DC, and feeds it into the grid. The blue arrow indicates the pilot signal transmitted by the GS to the satellite, which informs the satellite of its pointing direction. The red arrows indicate telemetry and demand information returned to the allocator.}
\label{fig:arch}
\end{figure*}

\section{Proposed System Model and Methodology}
We model 20 satellites in a Walker constellation that transmit microwave power at 2.45\,GHz to eight GSs distributed worldwide. The end-to-end architecture of the proposed system model for simulation is illustrated in Fig.~\ref{fig:arch}, comprising a constellation-level power allocator that supervises three pipeline segments: a \textit{Space~Segment} contains the satellite power chain, from PV generation through DC conditioning to the phased-array RF transmitter, a \textit{Free-Space~and~Atmospheric~Channel} carries the microwave beam from the satellite aperture to the rectenna, shaped by Goubau--Brown beam coupling and attenuated by ITU-R P.618 atmospheric losses, and a \textit{Ground Segment} converts the received RF back to DC at the rectenna and feeds the resulting power into the local grid.
The system jointly models orbital motion, the satellite power chain, beam physics, on-board battery behavior, and a power allocator that determines which satellite serves which station. The following subsection details each component as:

\subsection{Constellation Geometry and Orbital Dynamics}
The constellation comprises four orbital planes, each with five satellites, arranged in a Walker delta. All satellites operate at an altitude of 450\,km with a $53^\circ$ inclination. The same shell class is used by the lowest Starlink tier at 550\,km, and we chose a slightly lower altitude to keep the microwave beam footprint smaller on the ground. We propagated each satellite using Earth's $J_2$ oblateness, atmospheric drag modeled by an exponential density profile of the 1976 U.S.\ Standard Atmosphere~\cite {vallado2013fundamentals}, and solar radiation pressure, and we modeled eclipse events using a cylindrical Earth-shadow model. The orbital period is 93.5\,min, and the longest eclipse is about 36\,min when the orbital plane aligns with the Sun's direction. At each step, the simulation recomputes which GSs a satellite can see, using a $5^\circ$ minimum elevation mask.

\subsection{End-to-End Power Conversion Chain}
Each satellite carries a sun-tracking solar array of 5,000\,m\textsuperscript{2} populated with III--V triple-junction cells at a beginning-of-life efficiency ($\eta_\mathrm{PV}$) of 33\%. This aligns with the state of practice in space--grade photovoltaics reported by NASA~\cite{Rodgers_SBSP_2024}. Cell efficiency degrades at 1.0\%/year, a typical engineering estimate for III-V multijunction cells in the LEO radiation environment. The maximum power-point tracker introduces an additional 2--5\% loss ($\eta_\mathrm{MPPT}$) across the operating range. This loss is small relative to the harness and amplifier stages and is therefore absorbed into the manufacturer-rated $\eta_\mathrm{PV}$ for the headline conversion expression below. The conditioned DC output is routed through a harness with a 0.8\% resistive loss ($L_\mathrm{harness}$) to the satellite power bus, following Pucci \textit{et al.}~\cite{pucci_high_2024} for a 2.496\, GHz SBSP application. Thus, the cell-to-RF efficiency on the satellite side can be computed as:
\begin{equation}
    \eta_\mathrm{sat}
    = \eta_\mathrm{PV}\,(1 - L_\mathrm{harness})\,\eta_\mathrm{DC\text{-}RF}
    \approx 0.278.
    \label{eq:eta_sat}
\end{equation}

The heat from the DC--RF stage is added to the spacecraft's thermal node and rejected via dedicated radiator surfaces.

\subsection{Beam Pointing and Wireless Power Transmission}

We assume that each satellite's transmit aperture is 200\,m\,$\times$\,200\,m. The wavelength at 2.45\,GHz ($\lambda=c/f$) is 0.122\,m. The ground rectenna is 500\,m\,$\times$\,500\,m and the beam-collection efficiency follows the Goubau--Brown form~\cite{goubau_guided_1961, brown_history_1984}:

\begin{equation}
    \tau^2 = \frac{A_\mathrm{TX}\,A_\mathrm{RX}}{\lambda^2 R^2},
    \label{eq:tau_squared}
\end{equation}
\begin{equation}
    \eta_\mathrm{beam} = 1 - \exp(-\tau^2),
    \label{eq:eta_beam}
\end{equation}
where $R$ is the slant range. At zenith ($R = 450\,$km), $\tau^2 \approx 3.30$ and $\eta_\mathrm{beam} \approx 96\%$. At low-elevation passes ($R \approx 800\,$km), $\eta_\mathrm{beam}$ falls to about $65\%$. This change tracks the variation in beam efficiency as a satellite moves across a visibility window. We adopt a transmit-array pointing budget of $0.005^\circ$ ($87\,\mu$rad). Su \textit{et al.}~\cite{su_position_2024} report that ${\sim}100$\,m-aperture wireless-power-transfer (WPT) arrays produce beams about $0.005^\circ$ wide and recommend $0.0005^\circ$ pointing accuracy to keep centroid offsets below 5\% of the receiver diameter. Our LEO geometry permits a looser budget: at our 200\,m aperture and 450\,km nadir slant range, $0.005^\circ$ yields a beam-centroid offset of about 39\,m, comfortably within the 500\,m rectenna. Cheng \textit{et al.}~\cite{cheng_phased_2025} require $0.0023^\circ$ for a gigawatt-class GEO system, where the longer slant range amplifies any residual angular error. Atmospheric attenuation is modeled using ITU-R P.618 rain statistics~\cite{itu_r_p618_14} for each ground site. Received RF is rectified at 85\% peak and 72\% minimum efficiency, falling between the $>80\%$ single-element results and the $70$--$80\%$ array-level results reported in~\cite{sasaki_microwave_2013}.

\subsection{Onboard Power Bus and Battery Coupling}
Each satellite carries a 3.6\,kWh lithium-ion battery with Nickel-Manganese-Cobalt (NMC) chemistry, cycled to an 80\% depth of discharge. The discharge dynamics use a Peukert exponent of $k = 1.05$, which increases the effective capacity loss at the higher current levels observed during transmission. The open-circuit voltage is used to track the state of charge via a linear approximation. When the satellite is in sunlight, any surplus DC power not used for RF transmission is routed to the battery via a shunt regulator. When the satellite is in eclipse, a 500\,W housekeeping load draws power from the battery. Satellites with a state of charge below 20\% are removed from the allocator pool until they recharge.

\subsection{Constellation-Level Power Allocator}

At each step, the allocator examines all satellite-GS pairs and retains only those that satisfy all limits simultaneously, including minimum elevation, slant range, pointing tolerance, beam-collection threshold, and battery state. Three policies operate on the same feasible set as follows:
\begin{enumerate}
    \item \textbf{Greedy} --- This allocator sorts visible GSs by combined link-and-rectenna efficiency, then allocates the satellite's RF budget to the best link until the safety cap is reached, then to the next-best link until the budget runs out. Demand is ignored at this stage; a global $1.05\times$ over-delivery cap is applied at the end. The working analysis of the \textit{Greedy} is presented in Section~\ref{sec:2F}.
    \item \textbf{Proportional fair} --- This allocator performs water-filling based on channel quality. The allocation $p_k = \min(h_k/\mu, \text{cap}_k)$ maximizes $\sum_k \log(h_k p_k)$ subject to the satellite's budget, giving better channels a larger share until the safety cap binds.
    \item \textbf{Min-demand} --- This allocator visits GSs in order of best channel first and allocates the minimum power needed to exactly meet each station's demand. Move to the next station with whatever budget remains. Stations later in the order may receive nothing if the budget is exhausted.
\end{enumerate}
All three policies allow the allocator to split a satellite's RF budget among multiple GSs within a step, modeled as time-multiplexing over the 30\,s window. A global $1.05\times$ over-delivery cap is applied to the per-station total received DC after all satellites have been allocated, ensuring that no station exceeds its demand by more than 5\%.
\subsection{Working Analysis of the Greedy}\label{sec:2F}
At every $30$-second simulation step, the presented \textit{Greedy} allocator first builds the feasible set of GS satellite pairs by checking, for each visible pair, whether it meets the minimum elevation mask $(5^\circ)$, the slant-range and pointing-tolerance limits, the Goubau–Brown beam-collection threshold, and the satellite's battery state-of-charge floor (20\%). Any pair that fails any of these checks is discarded before allocation begins, so the greedy logic operates only on links that are physically and operationally viable at that instant.
\par Within this feasible set, the allocator selects visible GSs for each satellite based on their combined link-and-rectenna efficiency, which serves as a score that combines the beam-collection factor (which depends on slant range via the Goubau–Brown expression) and the rectenna's conversion efficiency at the corresponding incident power density. This ranking is purely channel-quality-driven; station demand plays no role at this stage.
\par The available RF power budget for each satellite is first allocated to the highest-ranked GS. Power is assigned to this best link up to the per-station safety cap ($1.05\times$ that station's demand). If the satellite still has budget remaining after reaching this cap, the allocator moves to the next-best-ranked GS and repeats the process, continuing down the ranked list until either the satellite's RF budget is exhausted or no further eligible GS remains in view. Because a satellite can see more than one station in a given window, its $30$-second budget can be split across multiple GSs. This is modeled as time-multiplexing within that step rather than true simultaneous multi-beam transmission.
This per-satellite allocation is computed independently for each satellite in the constellation at that timestep, and the resulting per-station totals are then aggregated across all satellites. Following power aggregation, a global over-delivery cap of $1.05\times$ is enforced at each GS, truncating the total received power if the combined multi-satellite contributions exceed this threshold.
\par The net effect, as presented in Section~\ref{sec:results}, is that the policy consistently favors the satellite–GS pair with the strongest instantaneous channel quality at each step. Stations with geometrically weaker passes (lower elevation, larger slant range, poorer beam coupling) are repeatedly outranked by competing links to other stations. This is why Cape Town and São Paulo received no service at all over the $24$-hour run. Their passes were never the top-ranked choice at any step when they were visible, so no RF budget ever reached them under this scoring rule.

\begin{table}[!t]
    \small
    \centering
    \caption{Simulation Parameters}
    \label{tab:sim_params}
    \renewcommand{\arraystretch}{1.1}
    \resizebox{0.8\linewidth}{!}{
    \begin{tabular}{@{} p{0.58\columnwidth} l @{}}
        \toprule
        \textbf{Parameter} & \textbf{Value} \\
        \midrule
        \multicolumn{2}{@{}c@{}}{\textbf{Constellation and Orbit}} \\
        Walker-delta layout & $4 \times 5 = 20$ satellites \\
        Altitude / Inclination & 450~km / $53^\circ$ \\
        Min elevation mask & $5^\circ$ \\
        \midrule
        \multicolumn{2}{@{}c@{}}{\textbf{Power Conversion}} \\
        PV array area & 5{,}000~m$^2$ \\
        PV BOL efficiency~\cite{Rodgers_SBSP_2024} & 33\% \\
        Annual degradation & 1.0\%/yr \\
        Harness loss & 0.8\% \\
        DC-to-RF efficiency~\cite{pucci_high_2024} & 85\% \\
        \midrule
        \multicolumn{2}{@{}c@{}}{\textbf{Wireless Power Transmission}} \\
        Frequency & 2.45~GHz \\
        TX aperture & 200~m $\times$ 200~m \\
        RX aperture (rectenna) & 500~m $\times$ 500~m \\
        Beam model~\cite{goubau_guided_1961, brown_history_1984} & Goubau--Brown \\
        Pointing budget~\cite{su_position_2024} & $0.005^\circ$ (87~$\mu$rad) \\
        Rectenna $\eta$ peak / min~\cite{sasaki_microwave_2013} & 85\% / 72\% \\
        Atmospheric model~\cite{itu_r_p618_14} & ITU-R~P.618 \\
        \midrule
        \multicolumn{2}{@{}c@{}}{\textbf{Battery and Bus}} \\
        Capacity / Chemistry & 3.6~kWh, Li-ion NMC \\
        Depth of discharge & 80\% \\
        Peukert exponent & $k = 1.05$ \\
        Min SOC for service & 20\% \\
        Eclipse housekeeping & 500~W \\
        Shunt regulator & on DC bus \\
        \midrule
        \multicolumn{2}{@{}c@{}}{\textbf{Allocator and Scheduling}} \\
        Simulation time step & 30~s \\
        Over-delivery cap & $1.05\times$ demand \\
        Beam splitting & disabled (1 RX / sat / step) \\
        Ground stations & 8 (0.8--2.0~MW, demo) \\
        \bottomrule
    \end{tabular}
    }
\end{table}

\section{Results}
\label{sec:results}
In this section, we evaluate the performance of the proposed 20-satellite Walker-delta SBSP constellation that delivers WPT via microwave at 2.45 GHz from a 450 km LEO. We consider four orbital planes, each with five satellites at an inclination of \ang{53}. Eight GSs equipped with rectennas are distributed globally, spanning latitudes from $-33.\ang{9}$ (Cape Town) to $55.\ang{8}$ (Moscow). Each satellite carries a 5000 m\textsuperscript{2} GaAs triple-junction solar array and a 200 m $\times$ 200 m transmit aperture targeting 500 m $\times$ 500 m rectennas. All results are obtained from a 24-hour simulation run using the greedy power allocator, with a 30-second time step (2880 steps in total). The rest of the simulation parameters are listed in Table~\ref{tab:sim_params} unless otherwise specified.
\begin{figure}[!t]
    \centering
    \includegraphics[width=1\linewidth]{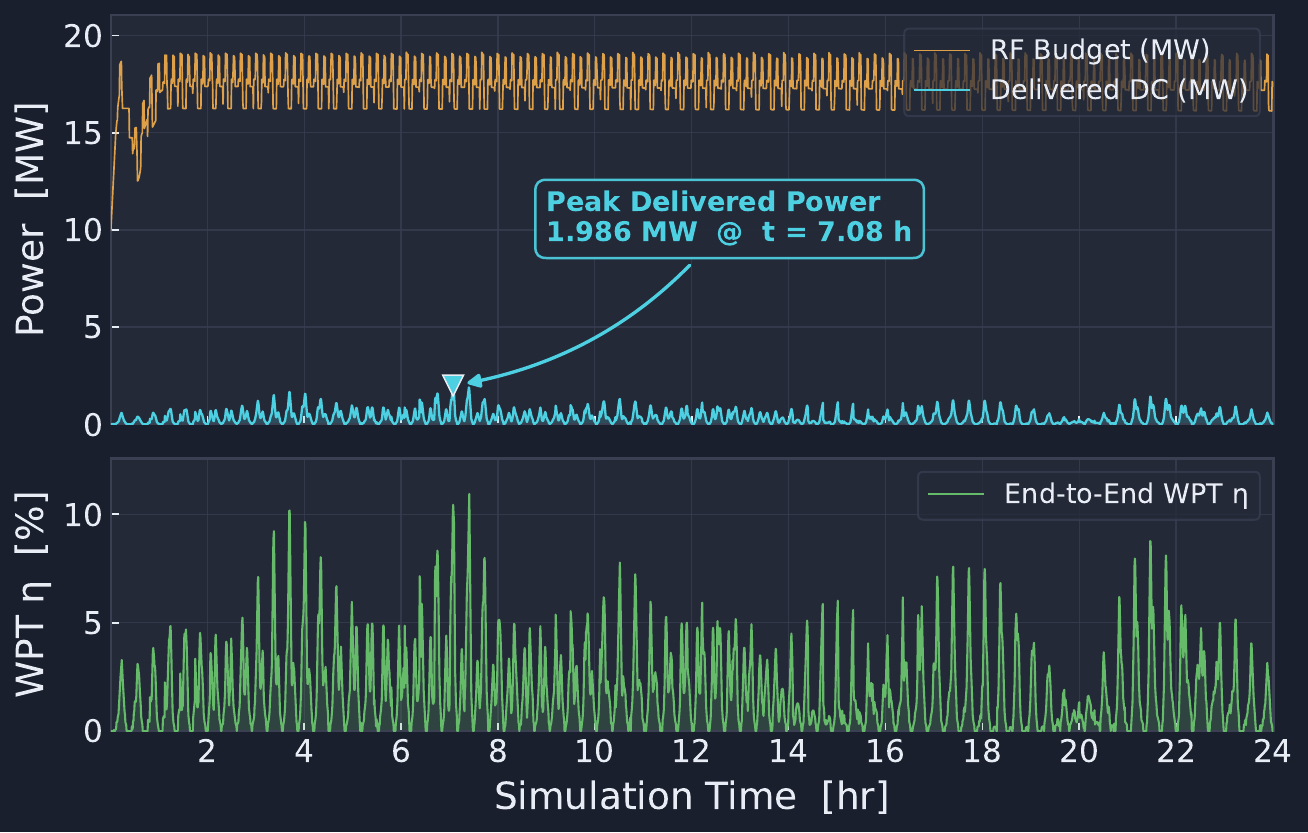}
    \caption{24-hour time series of constellation power flow under the greedy allocator policy. The top panel shows the constellation's instantaneous RF transmit budget (orange) and the total delivered DC power at the ground (cyan), both in megawatts. The bottom panel shows the corresponding end-to-end WPT efficiency. The triangular marker in the top panel indicates the peak delivered power of 1.986\,MW at $t = 7.08$\,h. The 16--18\,MW oscillation in the RF budget tracks the orbital eclipse cycle (period $\approx 93.5$\,min), with dips corresponding to eclipse passes during which only the on-board battery contributes.}
    \label{fig:2}
\end{figure}
The constellation's RF power budget and the total delivered DC power over 24 hours, along with end-to-end WPT efficiency, are shown in Fig.~\ref{fig:2}. The RF power budget stabilized at an oscillation between 16--18\,MW after the first orbital period. The constellation stabilizes the orbital coverage geometry within the first two hours and then enters a 92-min eclipse, during which solar power generation ceases, and only the battery-backed housekeeping bus remains active. Under a favorable satellite constellation geometry, the peak delivered DC power reaches 1.986\,MW around $t = 7.08$\,hr, as multiple satellites simultaneously achieve high-elevation passes over high-demand stations, yielding the best instantaneous beam-collection efficiency over the 24-hour window. Overall, delivered power is low (a few hundred kilowatts), reflecting the intermittent nature of LEO satellite coverage. The proposed greedy allocator efficiently serves the station with the strongest instantaneous link factor, leaving lower-priority or geometrically disadvantaged stations underserved for extended intervals.
The end-to-end WPT efficiency across the constellation is shown in Fig. 1(b). The efficiency peaks at nearly 11\% during the favorable pass at $t \approx 7$h and fluctuates between roughly 1\% and 10\% throughout the simulation. The low mean efficiency is primarily due to the beam-collection factor. At the nominal nadir slant range of approximately 490 km for a 450 km orbit, the maximum geometric collection efficiency is about 16\%, an intrinsic constraint of LEO geometry that cannot be overcome without significantly larger apertures or a higher orbit altitude.
\par
\begin{figure}[!t]
    \centering
    \includegraphics[width=1\linewidth]{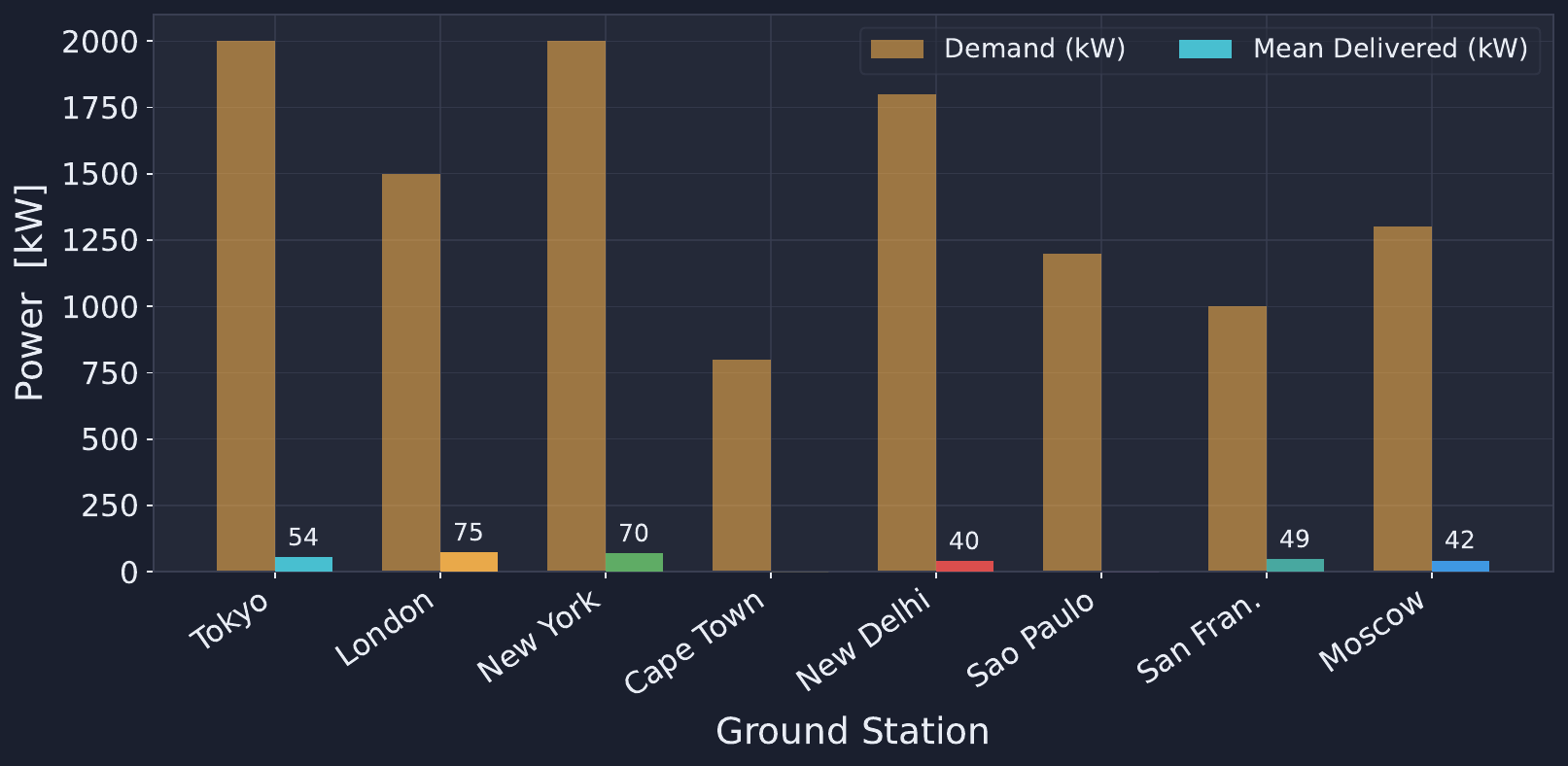}
    \caption{Per-station demand and 24-hour mean delivered DC power under the greedy allocator. Demand bars are shown in orange (assumed values, 0.8--2.0\,MW per site, demonstration scale). Mean delivered values at the served stations range from 40 to 75\,kW (numerical labels above each bar). Cape Town and S\~ao Paulo received no service during the run.}
    \label{fig:3}
\end{figure}

 Fig. \ref{fig:3} shows the mean delivered power and the power demand for each station in the constellation network. It can be observed that the six GSs, including Tokyo, London, New York, New Delhi, San Francisco, and Moscow, received an average power of 40--75\,kW, while Cape Town and São Paulo received no power. Both unserved GSs are in the Southern Hemisphere. The reason for this is discussed in Section~\ref{sec:discussion}. The shortfall relative to demand is the headline result of this work: at the constellation size and per-site demand that we modeled, the system delivers only a small fraction of the requested amount. The per-station floor and the served-versus-unserved split remain stable throughout the 24-hour window.

\begin{figure}[!b]
    \centering
    \includegraphics[width=1\linewidth]{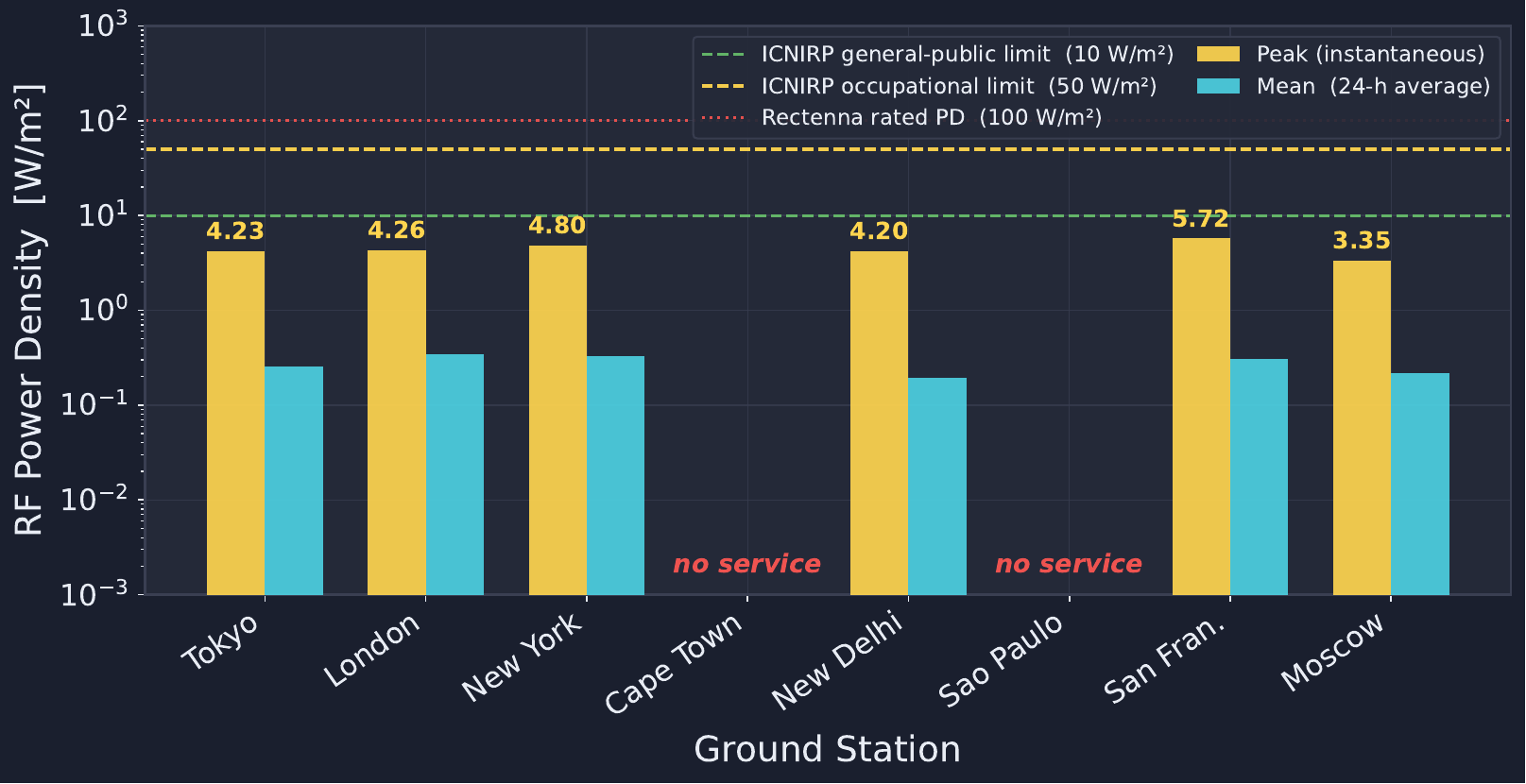}
    \caption{Per-station RF PD at the rectenna was measured over the 24-hour period. Yellow bars show peak instantaneous values (3.35--5.72\,W/m\textsuperscript{2} at the served stations); cyan bars show 24-hour means (0.19--0.34\,W/m\textsuperscript{2}). Three horizontal lines mark the relevant thresholds: the ICNIRP general-public exposure limit (10\,W/m\textsuperscript{2}), the ICNIRP occupational limit (50\,W/m\textsuperscript{2}), and the rectenna's rated PD (100\,W/m\textsuperscript{2}). The y-axis is logarithmic. Stations labeled \textbf{no service} received no transmissions during the run.}
    \label{fig:4}
\end{figure}
\begin{figure}[!t]
    \centering
    \includegraphics[width=1\linewidth]{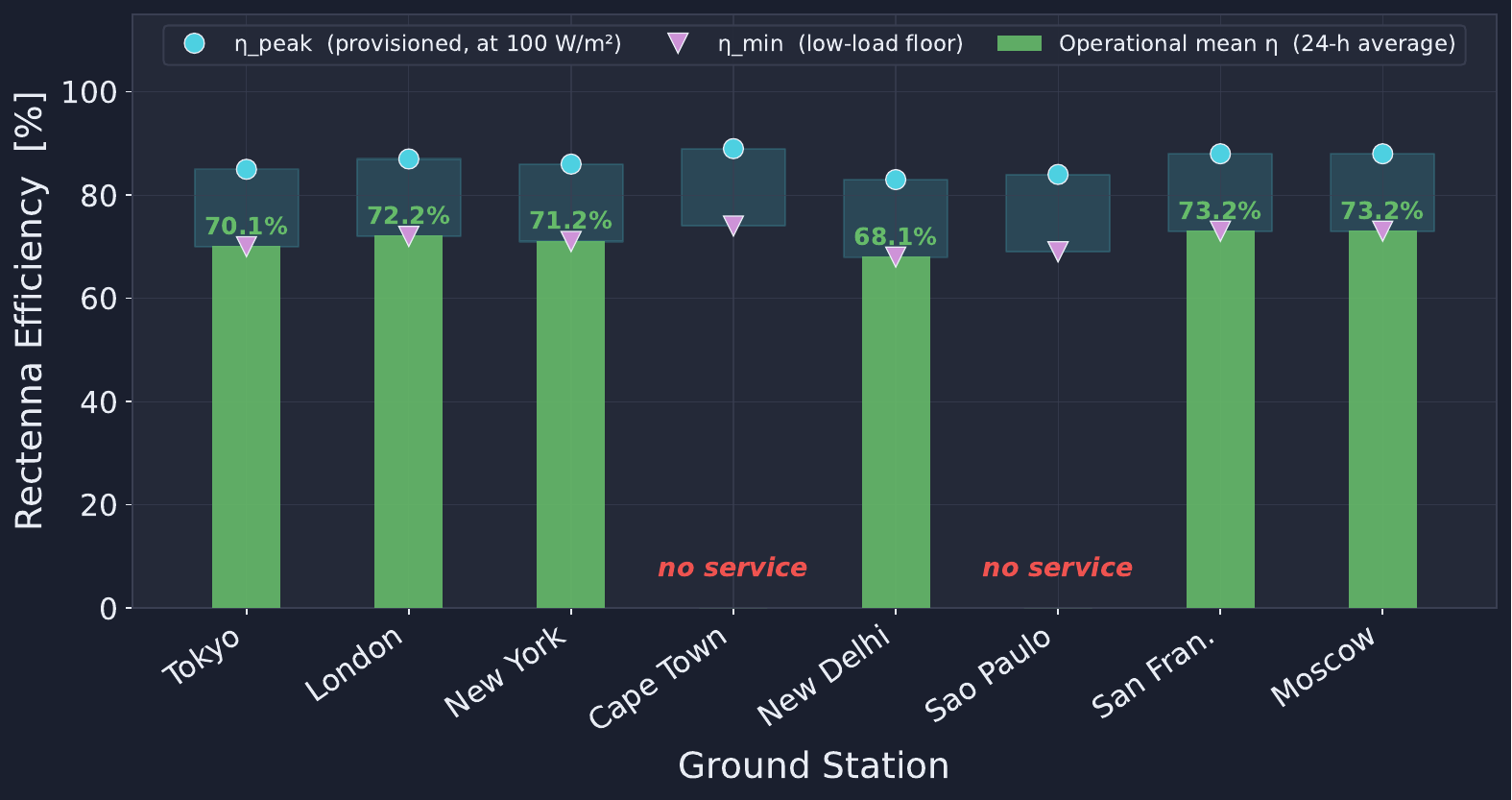}
    \caption{Per-station operational rectenna efficiency over the 24-hour run, plotted against each station's static design band. The shaded box at each station spans $\eta_{\min}$ (low-load floor, purple triangle) to $\eta_{\text{peak}}$ (provisioned at 100\,W/m\textsuperscript{2} rated PD, cyan circle). Green bars indicate the operational mean efficiency over actively served steps, with values of 68.1--73.2\,\% at the served stations. The operational means are close to $\eta_{\min}$ because the incident PD (Fig.~4) is two orders of magnitude below the 100\,W/m\textsuperscript{2} rated point.}
    \label{fig:5}
\end{figure}

Figs. \ref{fig:4} and \ref{fig:5} examine the corresponding incident PD and rectenna efficiency at each site. The instantaneous peak PD across the served stations falls within the 3.35--5.72\,W/m\textsuperscript{2} range, well below the ICNIRP general-public limit of 10\,W/m\textsuperscript{2} \cite{icnirp_2020_guidelines} and two orders of magnitude below the rectenna's rated PD of 100\,W/m\textsuperscript{2}. The mean PD over the 24-hour window is 0.19--0.34\,W/m\textsuperscript{2}. As a result, the operational rectenna efficiency tracks the low-load floor of the design band ($\eta_{\min}$) rather than the rated peak: operational means at the served stations are 68.1--73.2\,\%, close to the $\eta_{\min}$ markers and far below the $\eta_{\text{peak}}$ provisioned at 100\,W/m\textsuperscript{2}. For a constellation of this size, the rectenna is over-provisioned; sizing the rectifier to the expected mean PD would restore the efficiency margin.
\section{Discussion}
\label{sec:discussion}
The 24-hour run delivers a clear sizing message. The constellation, as modeled, yields an RF budget of 16--18\,MW but delivers only 40--75\,kW to each served station, roughly one-twentieth of the per-site demand. The shortfall does not stem from any single conversion stage. It results from the combination of \textbf{(i)} a sparse visibility pattern at LEO, where most satellites see no more than one GS at any given step, and \textbf{(ii)} a long power chain that compresses the budget by an order of magnitude before it reaches the grid. For a 20-satellite Walker-delta constellation at 450\,km altitude, the realistic per-site delivery is closer to 50--100\,kW than to a megawatt. This is the figure of merit a system designer should target at this scale, because any demand profile exceeding that level will be unmet.

Two of the eight GSs, Cape Town and São Paulo, received no service throughout the run. The greedy allocator ranks visible satellite-GS pairs by their combined link-and-rectenna efficiency and allocates each satellite's RF budget to the best pair first. The passes at these two sites were consistently outranked at every step where they were visible, leaving no budget for them. A demand-aware policy (i.e., min-demand or proportional fair with a guarantee floor) would spread the shortfall more evenly across all eight sites, even if the total delivered did not change. 

The operational rectenna efficiency in Fig. \ref{fig:5} lies at the bottom of the design band rather than at the peak. This is because the incident PD at the rectenna (Fig. \ref{fig:4}) is two orders of magnitude below the 100\,W/m\textsuperscript{2} rated value for which the rectifier was provisioned. For a constellation of this size, the rectenna is over-specified. Resizing the rectifier to a peak design point near 5\,W/m\textsuperscript{2} would shift its operating point into the high-efficiency region, recovering several percentage points of end-to-end efficiency without changing the satellite side. This is one practical takeaway from the run. Two limitations of the model are worth noting. The rectenna efficiency curve is a two-point design band ($\eta_{\text{peak}}/\eta_{\min}$) rather than a measured load-power curve, whereas a real rectifier's efficiency varies smoothly with input. The demand profile is also held constant at each site, whereas the real grid load follows a diurnal cycle. Both are reasonable simplifications for first-pass sizing, and we expect to relax them when extending the work to load chasing.

\section{Conclusion}
We presented a 24-hour, 30-second-step simulation of a Walker $4\times 5$ LEO SBSP constellation at an altitude of 450\,km, beaming 2.45\,GHz microwave power to eight GSs under a greedy allocation policy. The peak DC power delivered across the constellation reached 1.986\,MW, while the mean delivered power at the six served stations was 40--75\,kW, about one-twentieth of the per-site demand we set. Two of the stations, Cape Town and S\~ao Paulo, received no service during the run; their passes were outranked at every step by stronger links to other sites. The incident PD remained well below the ICNIRP general-public limit at every served site, with peak values of 3.35--5.72\,W/m\textsuperscript{2}. The operational rectenna efficiency tracked the floor of the design band rather than the peak because the incident PD was two orders of magnitude below the 100\,W/m\textsuperscript{2} rated point.

For a 20-satellite Walker LEO at this altitude, realistic per-site delivery is 50--100\,kW per station, and the rectenna should be sized to the actual incident PD rather than to a GEO-era-rated value. Future work will compare the three allocator policies under the same conditions and extend the simulation to time-varying demand profiles.

\bibliographystyle{IEEEtran}
\bibliography{references}

\end{document}